\documentclass[apj]{emulateapj}
\usepackage{graphicx}
\usepackage[toc,page]{appendix}
\usepackage{natbib}
\usepackage{color}
\usepackage{threeparttable}
\usepackage{hyperref}
\usepackage{breakurl}
\begin{document}
\slugcomment{To appear in the Astrophysical Journal}
\title{A Large Population of Massive Compact Post-Starburst Galaxies at $z>1$: 
Implications for the Size Evolution and Quenching Mechanism of Quiescent Galaxies}
\email{katherine.whitaker@yale.edu}
\author{Katherine E. Whitaker\altaffilmark{1}, Mariska Kriek\altaffilmark{2}, 
Pieter G. van Dokkum\altaffilmark{1}, Rachel Bezanson\altaffilmark{1}, 
Gabriel Brammer\altaffilmark{3}, Marijn Franx\altaffilmark{4}, Ivo Labb\'{e}\altaffilmark{4}}
\altaffiltext{1}{Department of Astronomy, Yale University, New Haven, CT 06511}
\altaffiltext{2}{Harvard-Smithsonian Center for Astrophysics, 60 Garden Street, Cambridge, MA 02138, USA}
\altaffiltext{3}{European Southern Observatory, Alonso de C\'{o}rdova 3107,Casilla 19001, Vitacura, Santiago, Chile}
\altaffiltext{4}{Sterrewacht Leiden, Leiden University, NL-2300 RA Leiden, The Netherlands}

\shortauthors{Whitaker et al.}
\shorttitle{Massive Compact Post-Starburst Galaxies at $z>1$}

\begin{abstract}
We study the growth of the red sequence through the number density
and structural evolution of a sample of young and old quiescent
galaxies at $0<z<2$.  The galaxies are selected 
from the NEWFIRM Medium-Band Survey (NMBS) in the Cosmic Evolution Survey 
(COSMOS) field.  We find a large population of massive young recently quenched (``post-starburst'')
galaxies at $z>1$ that are almost non-existent at $z<1$;
their number density is $5\times10^{-5}$ Mpc$^{-3}$ at $z$$=$2, 
whereas it is a factor of 10 less at $z=0.5$.
The observed number densities of young and old quiescent galaxies at $z>1$ are 
consistent with a simple model in which all old quiescent galaxies were
once identified as post-starburst galaxies.  We find that the overall population of 
quiescent galaxies have smaller sizes and slightly more elongated shapes
at higher redshift, in agreement with other recent studies.  
Interestingly, the most recently quenched galaxies at 
$1<z<2$ are not larger, and possibly even smaller, than older galaxies
at those redshifts.  This result is inconsistent with the idea that the 
evolution of the average size of quiescent galaxies is largely driven by
continuous transformations of larger, star-forming galaxies: in that case,
the youngest quiescent galaxies would also be the largest.  Instead,
mergers or other mechanisms appear to be required to explain the size growth
of quiescent galaxies from $z=2$ to the present.
\end{abstract}


\section{Introduction}
\label{sec:intro}

Galaxies with quiescent stellar populations form a well-defined relation in 
color-magnitude (color-mass) space known as the ``red sequence'', 
observed both in the local universe \citep[e.g.,][]{Kauffmann03,Baldry04}, 
as well as out to high redshift \citep[e.g.,][]{Franzetti07,
Cassata08,Kriek08b,Williams09,Brammer09,Whitaker10,Whitaker11}.
This red sequence is thought to grow over cosmic time by the 
continuous quenching and migration of star-forming galaxies, 
thereby changing the properties of these galaxies beyond simple 
passive evolution.

There is overwhelming evidence that the structures of these red sequence galaxies 
have undergone a rapid evolution since $z\sim2$ 
\citep[][etc.]{Daddi05,Cimatti08,vanDokkum08a}, where
galaxies of the same stellar mass have significantly smaller sizes in the past.  
Although it has been shown that red sequence galaxies will experience some growth through 
minor mergers and accretion \citep[e.g.,][]{Bezanson09, Naab09, Newman11}, the simplest 
explanation of this size growth is the continuous 
addition of recently quenched galaxies \citep[e.g.,][]{vanderWel09a}.  These galaxies 
that quench at later times are expected to have larger sizes and rounder shapes because the universe 
was less dense and therefore gas-rich, dissipative processes
were less efficient \citep[e.g.,][]{Robertson06,Khochfar06}.
The continuous transformation of these larger, star-forming galaxies at later times 
could be responsible for part of the size growth of the red sequence \citep[e.g.,][]{vanDokkum08a,
vanderWel09a, Hopkins09, Szomoru11}.

We can directly address the influence of transformations on the structural evolution 
of red sequence galaxies by looking to those galaxies that were most recently added.
In this scenario, these recently quenched galaxies will have larger sizes than older galaxies
on average at any given redshift. These young quiescent galaxies are 
easily identified when they pass through the ``post-starburst'' evolutionary phase
because their spectral energy distributions (SEDs) are dominated by A-type stars,
resulting in luminosity-weighted ages of $\sim1$ Gyr
\citep[e.g.,][]{LeBorgne06}. 
The structures of the population of young quiescent galaxies relative to older quiescent galaxies
provide crucial clues for the growth of the red sequence.

\begin{figure*}[t!]
\leavevmode
\centering
\plottwo{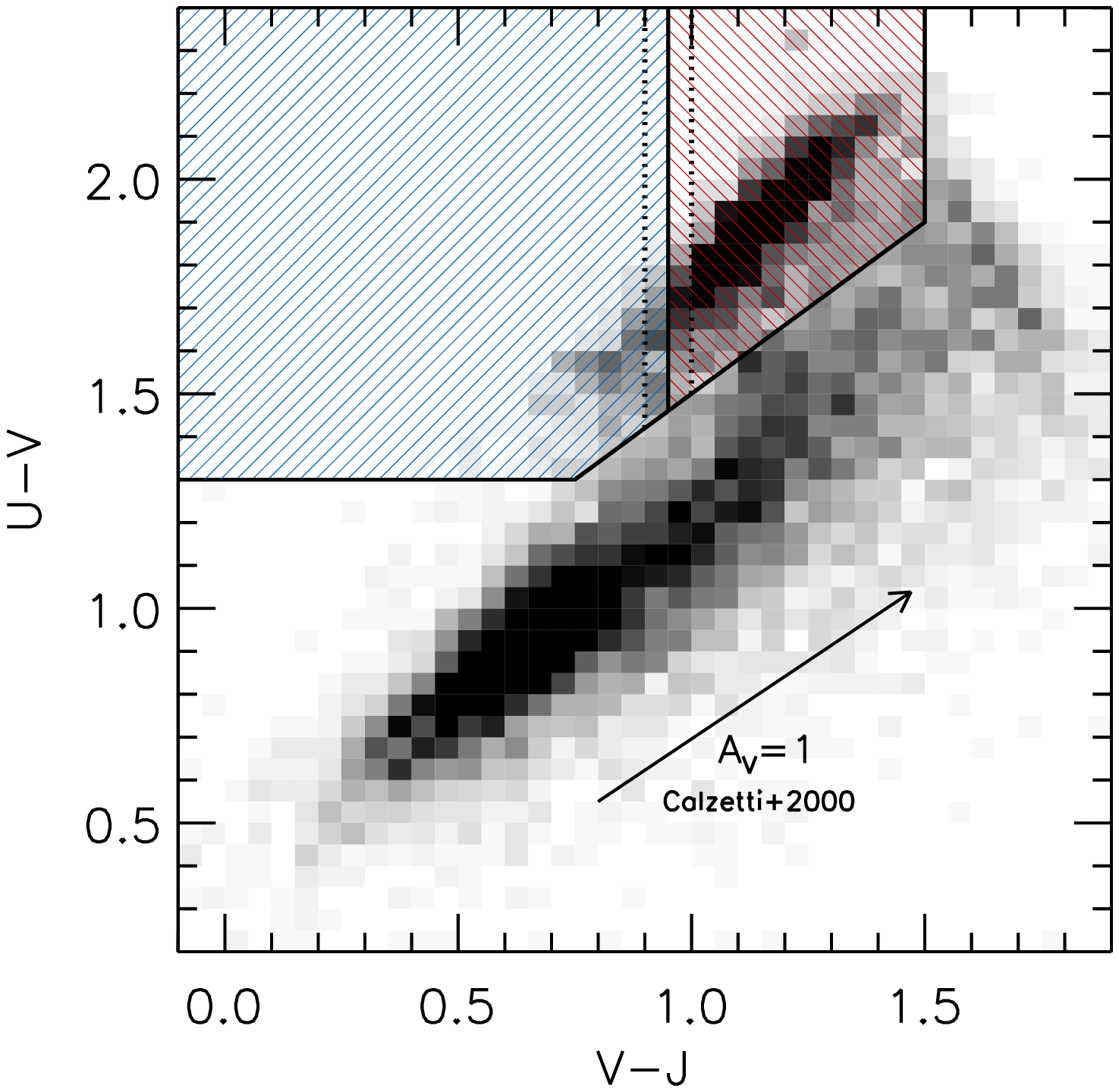}{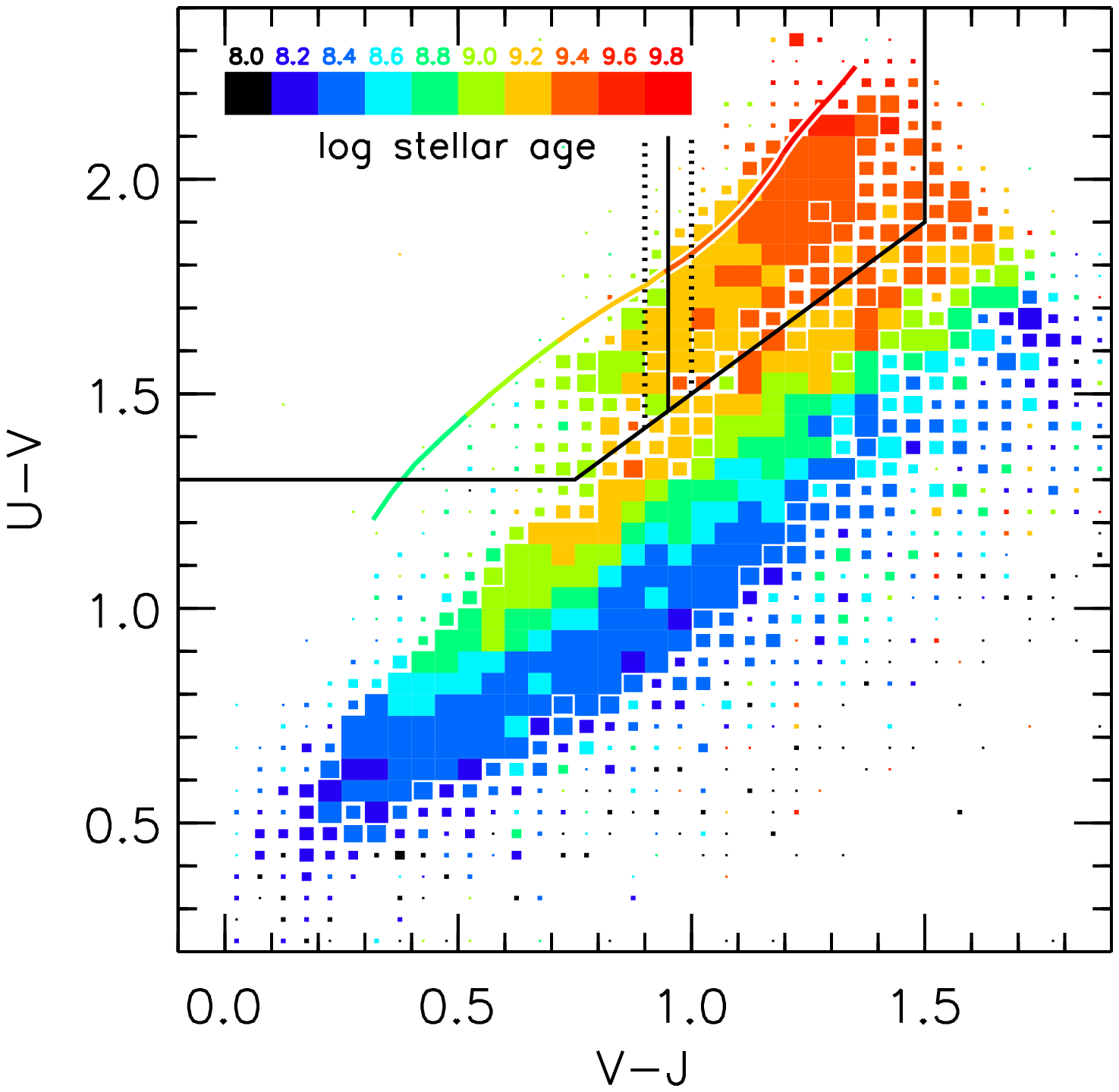}
\caption{(Left) Quiescent galaxies are selected based on their $U$--$V$ and $V$--$J$ colors, 
further divided into sub-samples of young (blue) and old (red) galaxies, where the greyscale 
represents the density of points. The separation of young and old quiescent galaxies is 
motivated by the age sequence of quiescent galaxies. (Right) The $U$--$V$ and $V$--$J$ 
colors of the entire NMBS parent sample, color-coded by the average best-fit 
age from \citet{BC03} stellar population synthesis models.  The size of the symbol reflects the 
number of galaxies in the bin.  The \citet{BC03} model track for a single burst stellar population 
with ages ranging from 0.5--10 Gyr is also shown. }
\label{fig:selection}
\end{figure*}

\begin{figure}[t!]
\leavevmode
\centering
\includegraphics[width=0.9\linewidth]{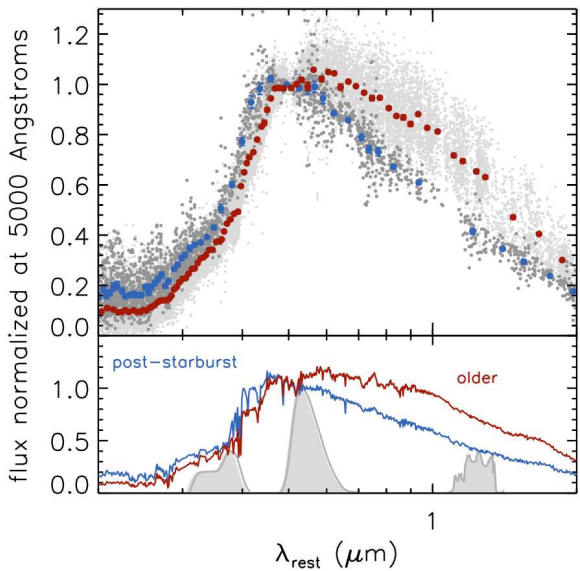}
\caption{Young and old quiescent galaxies are selected based on their $U$--$V$ and $V$--$J$ colors (see Figure~\ref{fig:selection}).
The composite rest-frame SED and running median of young (dark gray/blue) quiescent galaxies is distinct from that of 
old (light gray/red) quiescent galaxies (top panel).  The best-fit \citet{BC03} models and the $U$, $V$ and $J$
transmission curves are shown in the bottom panel.}
\label{fig:compositesed}
\end{figure}

Different evolutionary scenarios for the growth of the red sequence have been notoriously
difficult to test as recently quenched galaxies are hard to discriminate from older galaxies 
at high redshifts.  Owing to the excellent redshifts and accurate rest-frame colors of the 
NEWFIRM Medium-Band Survey \citep[NMBS;][]{Whitaker10,Whitaker11}, we
are now able to isolate samples of recently quenched galaxies. 
In this paper, we present the number density and structural evolution samples of
young and old quiescent galaxies out to $z=2.2$.  We further discuss mechanisms to 
produce the size growth of quiescent galaxies.

We assume a $\Lambda$CDM cosmology with $\Omega_{M}$=0.3, $\Omega_{\Lambda}$=0.7, 
and $H_{0}$=70 km s$^{-1}$ Mpc$^{-1}$ throughout the paper.  All magnitudes are 
given in the AB system.

\section{Data}
\label{sec:data}

Our sample of quiescent galaxies is drawn from the NMBS \citep{Whitaker11}.
This survey employs a new technique of using five medium-bandwidth near-infrared (NIR) filters to sample the
Balmer/4000\AA\ break from $1.5<z<3.5$ at a higher resolution than the standard
broadband NIR filters.  The combination of the medium-band NIR images with
deep optical medium and broadband photometry and IRAC imaging over 0.4 deg$^{2}$
results in accurate photometric redshifts
($\Delta z/(1+z)\lesssim 2$\%), rest-frame colors and stellar population
parameters \citep{Brammer09,Brammer11,vanDokkum10,Marchesini10,Whitaker10,Whitaker11,Kriek10,Kriek11,Wake11}.  
A comprehensive overview of the survey can be found in \citet{Whitaker11}.  
The stellar masses used in this work
are derived using FAST \citep{Kriek09a}, with a grid of \citet{BC03} 
models\footnote{We use the \citet{BC03} models following the results of 
\citet{Kriek10}, who find that the \citet{Maraston05} models do not fit 
the SEDs of young quiescent galaxies well, using the same dataset.} that assume a \citet{Chabrier}
IMF, solar metallicity, an exponentially declining star formation
histories and dust extinction following the \citet{Calzetti00} extinction law.

We use structural parameters derived from {\it Hubble Space Telescope} (HST) Advanced Camera for Surveys
(ACS) $F814W$-band ($I$-band)
imaging of the COSMOS field \citep{Scoville07} and $K_{S}$ broadband imaging from the
WIRCam Deep Survey (WIRDS; Bielby et al.,{\it in prep}).  
We refer to \citet{Bezanson11} for the details of the galaxy modeling.
The circularized effective radii and ellipticities presented in \S\ref{sec:structure} are determined using
S\'{e}rsic models convolved with a position-dependent point spread function (PSF) with GALFIT \citep{Peng02}.
The sizes are linearly interpolated between the observed ACS/WIRDS measurements to the rest-frame
wavelength of 8000$\mathrm{\AA}$, following \citet{Williams10}.  
Beyond $z>1.5$ we cannot measure ACS sizes for 8\% of the galaxies, as quiescent galaxies are 
inherently very faint in the rest-frame ultraviolet. Therefore we use the WIRDS size measurements alone.
We also include galaxies at $0.05<z<0.07$ selected from the
Sloan Digital Sky Survey (SDSS) Data Release 7 (DR7) \citep[see][]{Bezanson11}.

\section{Sample Selection}
\label{sec:selection}

Quiescent galaxies have strong Balmer/4000\AA\ breaks, characterized by red $U$--$V$ colors and blue 
$V$--$J$ colors relative to star-forming galaxies at the same $U$--$V$ color.  
With accurate rest-frame colors, it is possible to isolate ``true'' samples of quiescent galaxies 
(removing dusty, star-forming galaxies that contaminate the red sequence) using two rest-frame 
colors out to high redshifts \citep{Labbe05,Wuyts07,Williams09,Ilbert09,Brammer11,Whitaker11}.  
Our quiescent selection box is shown in Figure~\ref{fig:selection}
(defined by $U-V > 0.8\times(V-J)+0.7$, $U-V>1.3$ and $V-J<1.5$), 
with the larger NMBS parent sample shown in grey-scale. 
We select a mass-complete sample of 839 quiescent galaxies out to $z=2.2$ using a
fixed mass limit of $5\times10^{10}$ M$_{\odot}$.

\begin{figure}[t!]
\leavevmode
\centering
\includegraphics[scale=0.72]{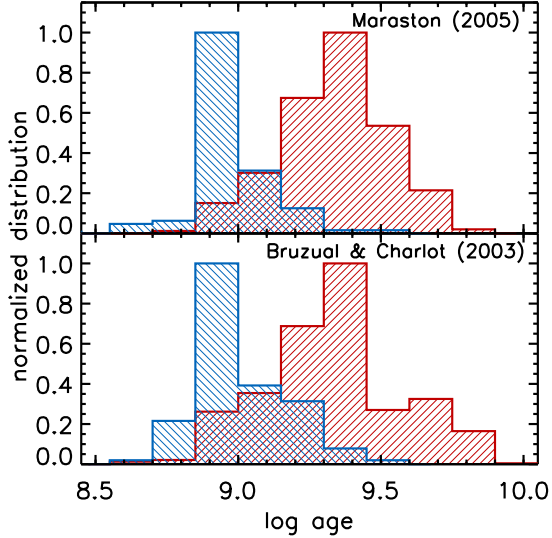}
\caption{The ages of young (blue) and older (red) quiescent galaxies from \citet{BC03} and 
\citet{Maraston05} models.  The two populations are separated independent of the models 
based on their $U$--$V$ and $V$--$J$ colors.}
\label{fig:ages}
\end{figure}

The stellar light of a recently quenched galaxy is dominated by A-type stars.
This results in spectral shapes distinct from older galaxies, with strong
Balmer-breaks and peaked SEDs compared to the 4000\AA\ break dominated spectrum
of older galaxies \citep[see also][]{Whitaker10}.
The youngest quiescent galaxies will have bluer $U$--$V$ and $V$--$J$ colors than
the older galaxies (blue, hashed region of Figure~\ref{fig:selection}).  
As the sequence in the $U$--$V$ and $V$--$J$ colors of quiescent galaxies is driven
by the ages of the stellar populations~\citep{Whitaker10}, 
we can separate 104 young quiescent galaxies with $V$--$J$ colors $<0.9$ from 
735 old quiescent galaxies.  
Through a visual inspection of the high resolution ACS imaging,
we find that 10\% and 7\% of the young and old samples are in pairs, respectively;
we remove these galaxies from the following analysis as their colors, redshifts and
stellar masses are unreliable because they were derived assuming they were a single galaxy.

The delineation between young and old is motivated by 
the average age of galaxies in $U$--$V$ and $V$--$J$ color space (see Figure~\ref{fig:selection}).
We show the \citet{BC03} model track for a single burst stellar population with ages ranging from 0.5--10 Gyr.
These model tracks do not include the effects of dust attenuation, which would shift the tracks to 
redder $U$--$V$ and $V$--$J$ colors.  Relatively small amounts of dust may reconcile the offsets between
the observed colors and the predictions from the models.  Regardless, the stellar population synthesis models
predict similar colors for the time period where a galaxy can be considered recently quenched. 
We note that dusty post-starburst galaxies may be missed with our adopted selection technique.

The difference between young and old quiescent
galaxies is clear when comparing their composite rest-frame SEDs (top panel in 
Figure~\ref{fig:compositesed}).  Although the young and old quiescent samples are separated
based on their rest-frame colors, which are determined independent of the stellar population synthesis 
models, the ages from both \citet{BC03} and \citet{Maraston05} models in Figure~\ref{fig:ages}
reflect the expected age distributions.  The median best-fit age of the young and old quiescent sample is
$1.0^{+0.4}_{-0.3}$ and $2.0^{+1.5}_{-0.9}$ Gyr,
where the error bars denote the $1\sigma$ spread within each sample rather than the (much smaller)
uncertainty in the mean age.
This young quiescent galaxy selection also agrees with predictions from 
simulations \citep[e.g.,][]{Wuyts09}.  Although the separation of young and old galaxies is 
somewhat arbitrary and there will be some crossover between the samples, the results presented herein 
are robust against changes in this threshold, as demonstrated in \S\ref{sec:numdens}.

We define our sample differently than spectroscopic samples of post-starburst galaxies, 
traditionally selected for strong H$\delta$
absorption with undetectable [O{\uppercase\expandafter{\romannumeral 2}}] emission lines that would
indicate ongoing star formation.  
We note that ``E+A'' (``K+A'') galaxies \citep[e.g.,][]{Dressler83,Couch87,Zabludoff96}
are a composite of an old stellar population with a recent burst,
whereas our SEDs are well-fit with a young burst alone.  As the stellar light is dominated by A-type
stars, this sample of post-starburst
galaxies could be described simply as ``A'' galaxies. 
The advantage of our selection is that we are able to isolate large, 
complete samples to high redshift that cannot be easily probed by spectroscopy. 

\section{Number Density Evolution}
\label{sec:numdens}

\begin{figure}[t!]
\leavevmode
\centering
\includegraphics[scale=0.45]{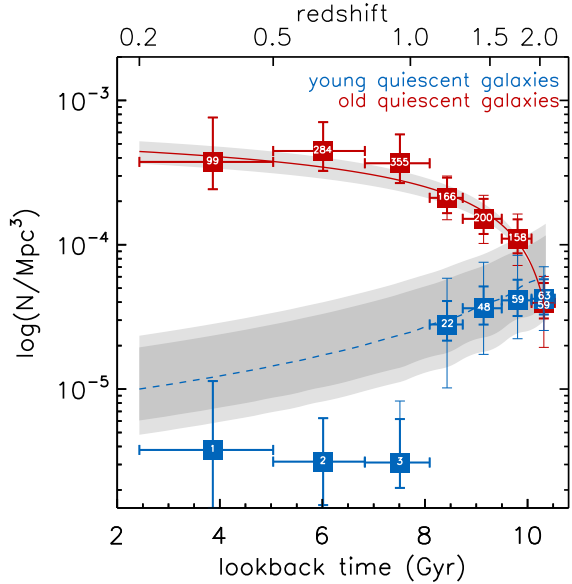}
\caption{The number density evolution of young (blue) and old (red) quiescent galaxies
reveals a dramatic rise in the population of recently quenched galaxies at $z>1$, with the number
of galaxies in each bin indicated.
The thicker errors include Poisson and cosmic variance errors computed 
following \citet{Somerville04}, added in quadrature. The thinner error bars span the full range
of Poisson and cosmic variance errors when changing
the threshold between young and old galaxies by $\pm0.05$ mag (indicated in Figure~\ref{fig:selection}). 
Assuming all galaxies pass through the
post-starburst phase, we predict the number of ``young'' quiescent galaxies (blue dashed line).}
\label{fig:numdens}
\end{figure}

In Figure~\ref{fig:numdens}, we present the number density evolution of the 839 
quiescent galaxies described in \S\ref{sec:selection}, in addition to 680 galaxies selected 
with the same criteria 
from the NMBS AEGIS field to reduce cosmic variance.  We see a dramatic rise in the number density
of massive young quiescent galaxies (blue) at $z>1$.  
At $z=2$, roughly half of all quiescent galaxies are relatively young,
which implies that we are most likely seeing the epoch where the red sequence is most-rapidly forming. 
This result is robust against changes in the threshold between young and old galaxies adopted in \S\ref{sec:selection}, 
as reflected with the error bars in Figure~\ref{fig:numdens}. 
If the star formation in these young quiescent galaxies at $z=1.5-2$ was quenched roughly 1 Gyr earlier,
the progenitors should be massive vigorously star-forming galaxies at $z=2-3$.   
The number density of older quiescent galaxies levels off at $z<1$, as the number density of young quiescent
galaxies rapidly drops.

Post-starburst galaxies are rare in the local universe: $<$1\% of galaxies at $z$=0 contain strong
Balmer lines and undetectable [O{\uppercase\expandafter{\romannumeral 2}}] emission lines \citep[e.g.,][]{Zabludoff96}.
The properties of relatively large samples of post-starburst galaxies have only been studied
out to $z\sim1$ \citep[e.g.,][]{Tran03,LeBorgne06,Wild09,Yan09,Vergani10}.  The observed fraction of massive galaxies with 
post-starburst spectral signatures increases from $<1\%$ at $z=0$ to $\sim10\%$ at $z=0.3-0.8$ \citep{Tran03}
upwards to $\sim20-50\%$ at $z=0.8-1.2$ \citep{LeBorgne06}.  Although we are not using the 
traditional spectroscopic selection for post-starburst galaxies, we are able to isolate samples of recently
quenched galaxies that would likely be classified as such if we had spectroscopic information available
at these high redshifts.  With this novel approach, we therefore push these studies to higher redshifts, finding consistent results.

These young quiescent galaxies will only stay in the post-starburst
phase for a relatively brief period, as the light from the A-stars that dominates the galaxies spectrum will
soon fade away.  Therefore, galaxies that we are catching in this phase will join the older
population roughly 0.5 Gyr later, barring any event that might trigger another burst of star 
formation.  

\begin{figure*}[t!]
\leavevmode
\centering
\includegraphics[scale=0.7]{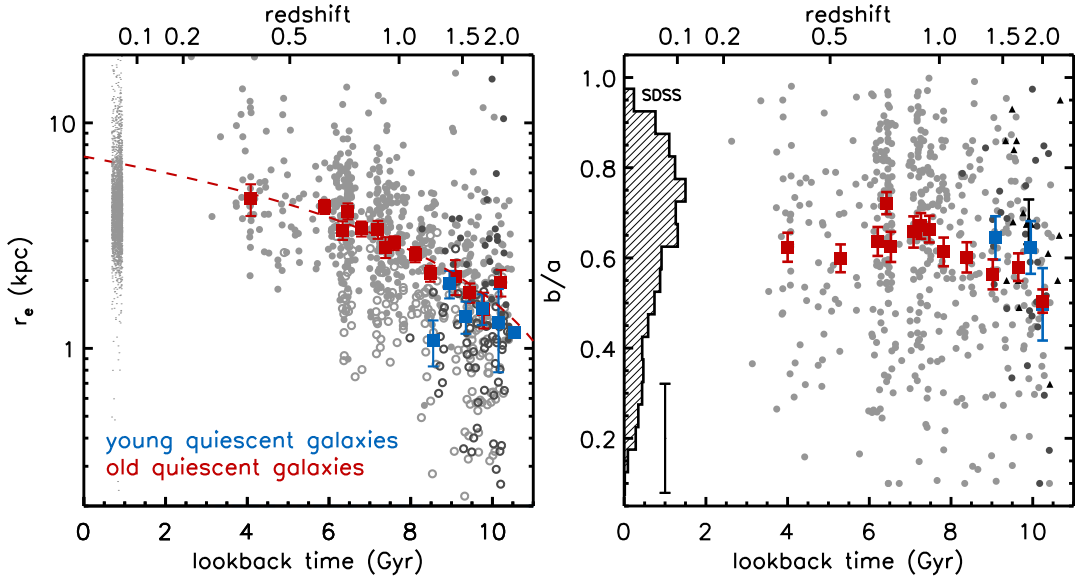}
\caption{The individual and median circularized effective radii (left) and axis ratios (right)
as a function of lookback time for young (dark grey/blue) and old (light grey/red) quiescent galaxies.
The sizes are normalized to a stellar mass of log M$_{\star}$/M$_{\odot}=11$.
The dashed line is the best-fit of $7.1-0.5t_{\mathrm{lb}}$ for old quiescent galaxies.
Open symbols are galaxies with $K_{S}$-band sizes less than 0.25$^{\prime\prime}$ 
and are not included in the axis ratio analysis.
The ellipticity measurements for 14 quiescent galaxies in the \cite{vanderWel11}
sample at $1.5<z<2.3$ are included (black triangles).  The lowest redshift data points (left panel) and histogram (right panel)
are massive galaxies at $0.05<z<0.07$ from the SDSS.  The error bar in the bottom left corner of the right
panel is the scatter when comparing WIRDS to higher resolution HST/WFC3 axis ratio measurements (see appendix).}
\label{fig:sizes}
\end{figure*}

Using the observed population of old quiescent galaxies, we model the growth of the massive end of the red sequence
by assuming that all quiescent galaxies pass through the post-starburst phase.
In this simple model, we fit the space density of older quiescent galaxies ($N_{\mathrm{old}}$) 
as function of redshift, finding a best-fit of:
\begin{equation}
N_{\mathrm{old}}(z)=0.0007-0.0002(1+z)
\label{eq:Nold}
\end{equation}

By assuming the timescale for the ``young'' phase ($\tau_{\mathrm{young}}$), we then 
predict the expected space density of young quiescent galaxies ($N_{\mathrm{young}}$), now
as a function of time:

\begin{equation}
N_{\mathrm{young}}\left(t\right) = N_{\mathrm{old}}\left(t+\tau_{\mathrm{young}}\right)-N_{\mathrm{old}}\left(t\right)
\label{eq:model}
\end{equation}

\noindent where $t$ is the age of the Universe.  In Figure~\ref{fig:numdens}, the best-fit to $N_{\mathrm{old}}$ 
from Equation~\ref{eq:Nold} and 1$\sigma$ uncertainty (red solid line/greyscale region) predicts the number of young quiescent galaxies 
(blue dashed line, $\tau_{\mathrm{young}}=0.5$ Gyr).
We explore a conservative range of $\tau_{\mathrm{young}}$ values from 0.3--1.0 Gyr (darker greyscale) 
for the full $1\sigma$ range (light greyscale), motivated by 
hydrodynamical simulations that result in the post-starburst phase~\citep{Snyder11}

This simple model is consistent with the observed number density evolution for young quiescent galaxies 
within the error bars.  The build
up of the massive end of the red sequence at $z>1$ can be explained by 
the fading of this large population of post-starburst galaxies.  Consequently, all quiescent 
galaxies at $z>1$ may have gone through the post-starburst phase.
Moreover, the drop in the number density of recently quenched galaxies at $z\sim1$ coincides with the 
epoch where the cosmic star formation rate density falls off \citep[e.g.,][]{Karim11}.

Because galaxies quenching at $z<1$ may not all pass through the post-starburst
phase, the most recent additions to the red sequence at late times will have more diverse star formation histories.
These results qualitatively agree with \citet{Wild09}, who find that galaxies passing through
this phase account for about half of the growth rate of the red sequence at $0.5<z<1$.
The number density of red sequence galaxies will also increase with (minor) mergers, as 
galaxies that fall below the mass limit may enter the sample at later times.  
This may be the dominant mechanism for growth of the red sequence at $z<1$.

\section{Structural Evolution}
\label{sec:structure}

We compare the structural properties of 685 old and 91 young quiescent galaxies, a
sample that is a factor of a few larger and extends to higher redshifts than previously probed.
In Figure~\ref{fig:sizes}, we measure a smooth increase in the median size of red sequence galaxies 
since $z=2$, finding consistent results with other similar 
studies \citep[e.g.,][and many more]{Daddi05,Trujillo07,vanDokkum10,Williams10}.  
The sizes are normalized to a stellar mass of log M$_{\star}$/M$_{\odot}=11$, using the size-mass relation 
from \citet{Shen03}. There does not appear to be a significant difference in the size growth of young
and old quiescent galaxies.  The measured size ratio of old to young quiescent galaxies is 1.4$\pm$0.5 at
$1.2<z<2.2$.

In the right panel of Figure~\ref{fig:sizes}, we find that the axis ratios of quiescent galaxies 
are becoming slightly flatter (possibly more disky) towards earlier times in the universe.   
The distribution of $I$-band ellipticities for the most massive (bulge-dominated) galaxies
in the SDSS at $0.05<z<0.07$ extends almost the full range of values, peaking near b/a=0.7.  
The observed trends are consistent with a mild evolution towards rounder shapes 
at lower redshifts, similar to the results by \citet{vanderWel11}.  As these structural parameters are 
derived from ground-based imaging, we compare the measurements for a subset of the sample with high-resolution
space-based imaging in the appendix.
We find that the space-based imaging yields similar measurements of the structural parameters to 
the ground-based measurements presented in this paper.

The structural evolution of young and old galaxies appears to be similar, but to do a more thorough analysis we 
stack the ACS $F814W$-band ($I$-band) and the WIRDS $K_{S}$-band images.
We create the stacked images by adding normalized, masked images of the individual galaxies at $1.5<z<2.0$. 
The postage stamps of the individual objects are normalized in two separate methods: 
all individual galaxies are ``mass''
normalized to a stellar mass of 10$^{11}$ M$_{\odot}$ and a redshift of $z=1.5$, and ``flux'' 
normalized to the average total flux of the entire quiescent sample before stacking.  
The mass normalization will be sensitive to any intrinsic brightness differences, whereas 
the flux normalization will emphasize any size differences.

\begin{figure}[t!]
\leavevmode
\centering
\includegraphics[scale=0.76]{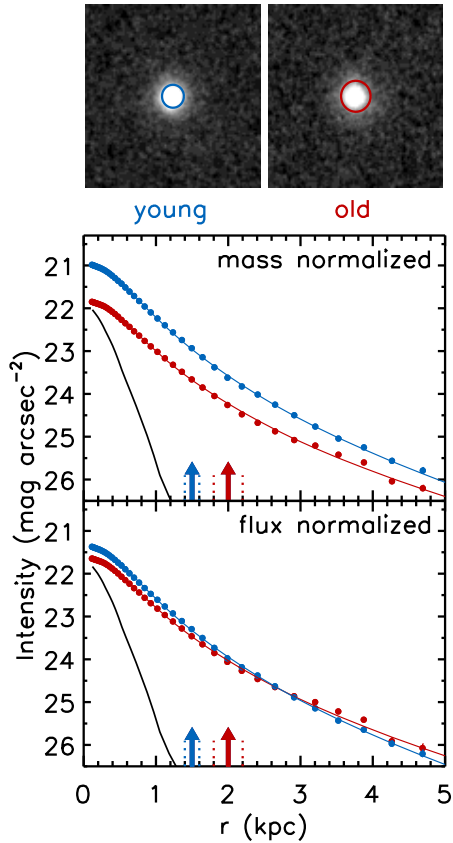}
\caption{The stacked ACS $I$-band postage stamps and surface brightness profiles
of young (blue) and old (red) quiescent galaxies at $1.5<z<2.0$, and the PSF (black).
The individual galaxies are ``mass'' (top) or ``flux'' normalized (bottom) before stacking. 
The error bars are generally smaller than the symbol size. 
The younger quiescent galaxies are more centrally concentrated with marginally
smaller sizes (arrows/circles). }
\label{fig:stacks}
\end{figure}

The flux normalized observed ACS $I$-band stacks are shown in Figure~\ref{fig:stacks}.
We show the surface brightness profiles
of the stacks for both normalization methods below the postage stamps,
with arrows indicating the mean sizes determined using GALFIT (see \S\ref{sec:data}).
We determine the mean using bi-square weighting and estimate the robust dispersion (resistant to outliers) 
for the effective radius and S\'{e}rsic index distributions from 50 bootstrapped stacks.
We present the structural parameters measured from this stacking analysis for both the $I$-band and $K_{S}$-band in
Table~\ref{tab:sbp}.  The surface brightness profiles are very similar within 5 kpc
for all quiescent galaxies, but we do detect subtle
differences.  On average, the younger quiescent galaxies are more centrally concentrated 
and may have somewhat smaller sizes, assuming that mass follows light.

As the mass normalization takes into account any brightness differences due to redshift, the remaining difference
of $\sim1$ mag between the central surface brightness of young and old galaxies is intrinsic.
The surface brightness profiles of the flux
normalized stacks begin to diverge at radii $>2$ kpc, with measured sizes in agreement with the mass
normalized stacks.

\section{Summary and Conclusions}
\label{sec:summary}

We study the growth of the red sequence through the number density
and structural evolution of a sample of young and old quiescent
galaxies selected from the NEWFIRM Medium-Band Survey.  Due
to the higher spectral resolution photometry of the NMBS, we can now --- for the
first time --- study the properties of large samples of massive young quiescent galaxies
at $z>1$.  We do not see many massive young galaxies that
have recently quenched their star formation locally or even at low to 
intermediate redshifts.  In fact, we only begin to see a dramatic rise
in the population of recently quenched post-starburst galaxies at $z>1$, finding an order of 
magnitude increase in their number densities between $z=0.5$ and $z=2$.  
The observed number densities of young and old quiescent galaxies are consistent with
a simple model where all quiescent galaxies at $z>1$ have passed 
through the post-starburst evolutionary phase.  

As the simple model used in this paper does not
take into account growth due to mergers, we repeat the analysis selecting galaxies 
(both quiescent and star-forming) at a constant number density of $2\times10^{-4}$ Mpc$^{-3}$,
following \citet{vanDokkum10}.  When taking into account mass growth on the red
sequence by both mergers and the transformation of star-forming galaxies, we find an even
more pronounced rise in the number density of young quiescent galaxies from $<10^{-6}$ Mpc$^{-3}$
at $z<1$ to $4\times10^{-5}$ Mpc$^{-3}$ at $1.5<z<2$.

Somewhat surprisingly, the younger quiescent galaxies are not larger, and 
perhaps even somewhat smaller, than the older galaxies at a fixed redshift. 
The compact nature of recently quenched galaxies has implications for both 
the size growth of the red sequence and the mechanism by which galaxies quench their star formation.  

If there is a faint extended halo around post-starburst galaxies at large radii that we are not
resolving, the measured size will underestimate the true radial distribution of the stellar light \citep[e.g.,][]{Hopkins09}.
Although we measure more centrally concentrated surface brightness profiles for younger quiescent galaxies,
there does not seem to exist a substantial fraction of stellar light in the outskirts that
we are missing.  The surface brightness profiles from the stacked images of both young and
old quiescent galaxies are similar out to 5 kpc.  With a careful analysis of quiescent 
galaxies, \citet{Szomoru10, Szomoru11} similarly rule out the existence of a faint
extended envelope around the observed galaxies down to the surface brightness limit of $H\sim28$
mag arcsec$^{-2}$.  The size growth of quiescent galaxies does not seem to be a consequence of
how we measure galaxy sizes.

The apparent lack of a size difference may not be intrinsic if the star formation history varies over the galaxy.
In the simplest scenario where no major structural evolution occurs, massive blue galaxies may have a 
younger population in the central regions. This population may not necessarily constitute a  
substantial fraction of the galaxy's stellar mass, but it is very luminous.  This 
would lead to a post-starburst galaxy that is both bright and blue in the center but older 
in the outskirts.  Aging of the stellar populations will lessen the differences in 
brightness between the center and the rest of the galaxy and lead to an increasing size.  
This scenario therefore predicts gradients in the stellar populations of quiescent galaxies, 
with an older stellar population in a faint extended halo \citep[e.g.,][]{Pracy11}.  
However, both \citet{vanDokkumBrammer10} and \citet{Guo11} have found the opposite, the inner regions
of quiescent galaxies at $z\sim2$ have redder rest-frame colors than the outer regions.
We find no strong evidence for color gradients in this data set, as the ratio of the 
mean size of the old and young quiescent galaxies is approximately independent of the passband.

The small size of recently quenched post-starburst galaxies directly contradicts 
the theory that the average size of quiescent galaxies increases over time due to the 
continuous quenching and migration of larger, star-forming galaxies \citep[e.g.,][]{vanderWel09b}.
Rather, as both young and old galaxies appear to have similar sizes, the size growth of quiescent
galaxies from $z=2$ to the present is likely the result of minor mergers and accretion of smaller galaxies. 

The compact nature of these massive post-starburst galaxies is puzzling in the context of a coherent 
evolutionary picture.  Similarly massive star-forming galaxies at the same epoch are likely 
experiencing their last burst of star formation and will shortly quench and migrate to the red sequence.  
Without invoking quenching mechanisms, the structures of star-forming and quiescent galaxies must be 
connected in some way.  However, a simple size-mass relation is inadequate to describe the size 
variations of the entire galaxy population at $z=2$.  At all epochs, star-forming galaxies are 
the largest among galaxies of a given stellar mass \citep[e.g.,][]{Williams10, Wuyts11b}.  
Furthermore, about half of massive star-forming galaxies at $z\sim2$ are observed to have 
clumpy structures \citep[e.g.,][]{ForsterSchreiber11,Swinbank10}. For these star-forming galaxies to be the progenitors of 
small quiescent galaxies, a compact core must already be present if the star formation quenches 
due only to the exhaustion of the gas reservoir.  Albeit for a small sample, \citet{Kriek09b} 
find that at least half of star-forming galaxies at $z\sim2$ require major structural 
changes beyond simple passive evolution to transform into these compact quenched systems.  
The process that makes these galaxies small may also quench the star formation.  Alternatively,
the samples of star-forming galaxies might be incomplete.

Understanding the structural evolution of galaxies may hold the clues to why they transition 
from the active to passive phase.  Gas exhaustion may not be the sole driver of galaxies 
migrating to the red sequence, as it is during this epoch that the cosmic star formation 
rate density peaks \citep[e.g.,][]{Hopkins06}, and galaxy halos are predicted to have large 
accretion rates \citep[e.g.,][]{Dekel09}.  Moreover, the structures of galaxies are observed 
to be related to the stellar populations, implying a causal connection between the quenching 
and the morphological transition.

Among proposed quenching mechanisms that result in (post-)starbursts are gas-rich major mergers 
\citep[e.g.,][]{Mihos94} after which point the available gas is exhausted from supernovae winds or 
Active Galactic Nuclei (AGN) feedback \citep[e.g.,][]{Croton06,PHopkins06}, shock heating in massive 
halos \citep{Birnboim07,Yan09}, or cold streams resulting in clump migration and coalescence \citep[e.g.,][]{Dekel09b}.  
The transformation mechanism responsible may drive cold gas to the central disk of the 
halo to trigger a starburst, thereby changing the structure of the galaxy and subsequently 
quenching star formation.  It is clear that more information about the stellar populations of 
massive star-forming and quiescent galaxies is necessary before we can begin to understand the 
connection between the active and passive phases of galaxies.  Future detailed studies of the 
environment and morphologies of both post-starburst and star-forming galaxies at high redshift 
may be able to disentangle the relative importance of these mechanisms.

\begin{acknowledgements}
We thank the anonymous referee for useful comments and a
careful reading of the paper.
We thank the NMBS collaboration for their contribution to this work and 
the COSMOS and AEGIS teams for the release of high quality
multi-wavelength data sets to the community.
Support from NSF grant AST-0807974 and NASA grant NNX11AB08G is gratefully acknowledged.
\end{acknowledgements}

\facility{\emph{facilities}: Mayall (NEWFIRM)}

\addcontentsline{toc}{chapter}{\numberline {}{\sc References}}
\bibliography{master}

\begin{table*}[h!]
  \caption{Structural Parameters of Quiescent Galaxies at $1.5<z<2.0$}
  \centering
  \begin{threeparttable}
    \begin{tabular}{lcccc}
      \hline
      \hline
        & \multicolumn{2}{c}{mass normalization} & \multicolumn{2}{c}{flux normalization} \\
      \cline{2-5}
       & $r_{e}$ (kpc) & n & $r_{e}$ (kpc) & n \\
      \hline
      Young Quiescent Galaxies (N$_{\mathrm{gal}}$ = 51) \\
      \hline
      ACS $I$-band & 1.5$\pm$0.1 & 2.5$\pm$0.2 & 1.5$\pm$0.1  & 2.4$\pm$0.2 \\
      WIRDS $K_{S}$-band  & 1.3$\pm$0.1 & 2.6$\pm$0.6 & 1.4$\pm$0.2  & 2.6$\pm$0.7 \\
      \hline
      Old Quiescent Galaxies  (N$_{\mathrm{gal}}$ = 81) \\
      \hline
      ACS $I$-band & 2.0$\pm$0.2  & 2.9$\pm$0.2 & 2.0$\pm$0.2  & 2.8$\pm$0.2 \\
      WIRDS $K_{S}$-band  & 2.1$\pm$0.5  & 9.1$\pm$6.0 & 2.2$\pm$0.4  &  8.4$\pm$5.8 \\
      \hline
    \end{tabular}
  \end{threeparttable}
  \label{tab:sbp}
\end{table*}

\newpage
\section*{Appendix. Derivation of Structural Properties}
\vspace{0.2cm}

\begin{figure*}[t!]
\leavevmode
\centering
\plottwo{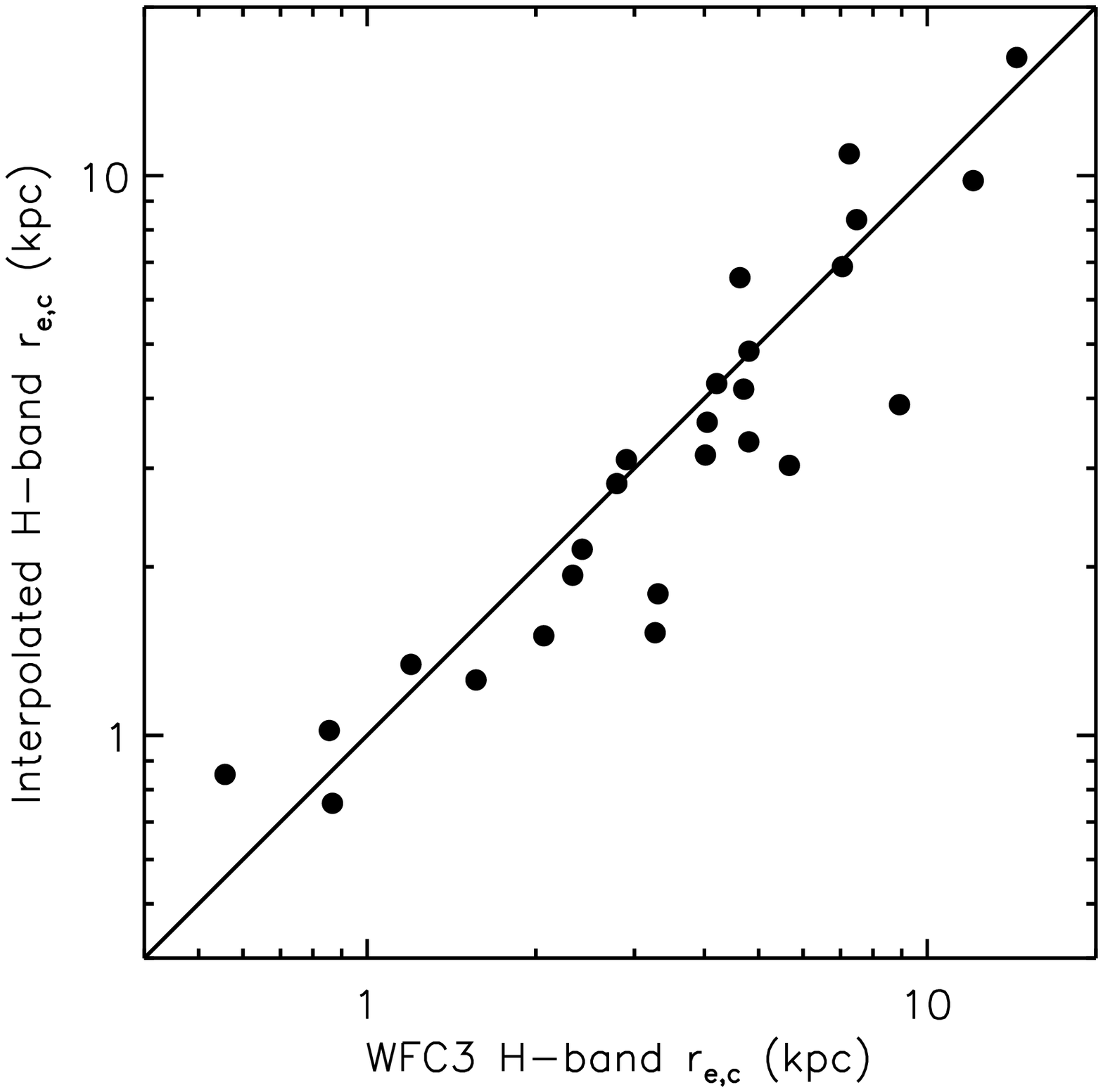}{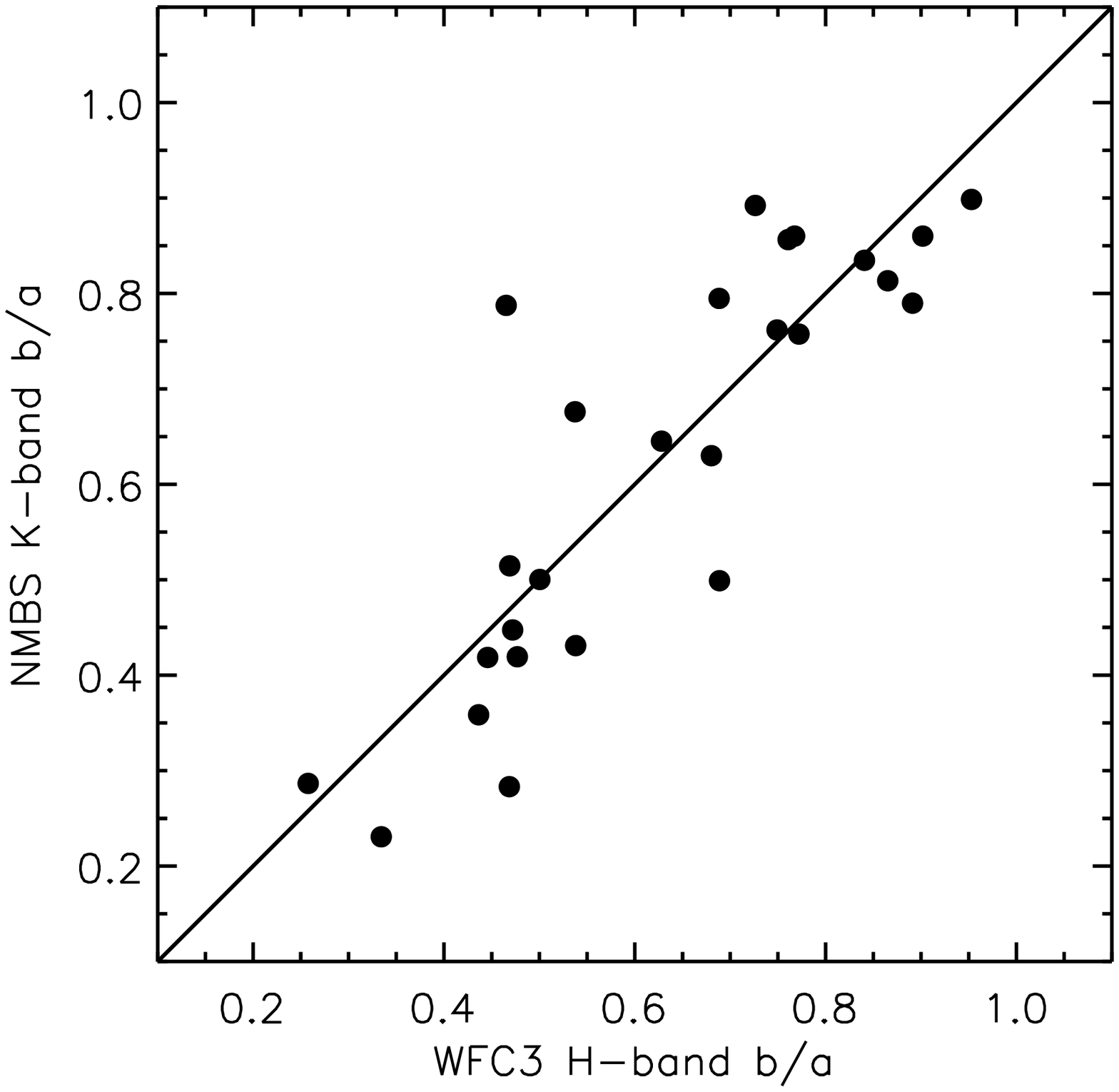}
\caption{(Left) The WFC3 $F140W$-band circularized and mass normalized sizes compared to ACS/WIRDS, interpolated at the central wavelength 
of the WFC3 $F140W$ filter. (Right) The WFC3 and WIRDS axis ratio measurements.  The structural parameters based on ground-based
imaging are consistent with the results using higher-resolution space-based imaging.}
\label{fig:wfc3}
\end{figure*}

In this paper we derive structural properties of quiescent galaxies based on $K_{S}$-band 
ground-based imaging.  Given the small angular sizes of the galaxies at the highest redshifts,
we compare our size and axis ratio measurements to high-resolution HST/WFC3 $F140W$-band ($H$-band) 
imaging for a subset of the galaxies to ensure that the measurements are robust.  
This subset of 26 quiescent galaxies have redshifts ranging from $z=0.5-2$, with an average value of $z=1.1$.
In Figure~\ref{fig:wfc3},
we find that ground-based interpolated $H$-band sizes are on average 0.05$\pm$0.03 dex larger, with a scatter of 0.14 dex.  
There is a scatter of 0.1 between the ground-based $K_{S}$-band and space-based $H$-band 
axis ratio measurements, with no systematic
offsets.  In general, the space-based imaging yields similar measurements of the structural parameters to the 
ground-based measurements presented in this paper.  Although there is a larger uncertainty in the structural
parameters when using ground-based measurements,  we assert that the general trends would be unchanged.

We find that the results of this paper are not very dependent on the linear interpolation of the rest-frame sizes, 
as the sizes in the $K_{S}$-band are generally not very different from the sizes in the $I$-band.  
More to the point, the ratio of the mean size of 
the old galaxies to the mean size of the young galaxies at $1.5<z<2$ is approximately independent of the passband: 
1.3$\pm$0.2 in the $I$-band and 1.6$\pm$0.4 in the $K_{S}$-band.
\end{document}